\documentclass[12pt,a4paper]{scrartcl}
\usepackage[T1]{fontenc}
\usepackage{newtxtext,newtxmath}
\usepackage{amsmath,amssymb}
\usepackage{bbm}
\numberwithin{equation}{section}
\usepackage[title]{appendix}

\usepackage{geometry}
\usepackage{braket}
\usepackage[pdftex]{graphicx}
\title{Virasoro algebra in $K$-space}
\author{Syoji Zeze\footnote{ztaro21@gmail.com} \\
Yokote Seiryo Gakuin High School\\
147-1 Maeda, Osawa, Yokote, 013-0041 Japan}
\date{}
\begin{document}
\maketitle 
\begin{abstract}
We introduce a novel representation of Virasoro algebra in open string field theory. Elements of the algebra are vector fields on the $K$-space, where $K$ is the string field that generates a world sheet strip in sliver frame.   The generators introduce a global symmetry of open string field theory.   We also derive a representation of Virasoro algebra for nontrivial open string background described by the formal pure gauge solution. Generators for 0, 1 and 2 D-branes are explicitly studied and different behaviors around $K=0$ are observed.
\end{abstract}

\section{Introduction}
One of an important subject in Open String Field Theory (OSFT) has been classical solutions in universal sector.  Solutions for the tachyon vacuum and D-branes belong to this sector.  Recently developments in this subject have reveled the nature of this sector both numerically~\cite{Kudrna:2018mxa} and analytically~\cite{Hata:2019dwu}.
Universal sector plays fundamental role in OSFT since
it is independent of a choice of open or closed string background~\cite{Sen:1999xm}.  

This paper is motivated by the recent development of Sine Square Deformation (SSD)~\cite{2009PThPh.122..953G} based technique for classical solutions of OSFT in universal sector~\cite{Kishimoto:2018ekq,Zeze:2018lao}.  SSD is a certain kind of deformation of one dimensional open spin lattice system in which the ground state spectrum coincides with that of the periodic system~\cite{2009PThPh.122..953G}.    The coincidence between ground state spectra has been explained by a Hamiltonian of two dimensional CFT~\cite{Katsura:2011ss}.  Quite remarkably, Ishibashi and Tada have derived a continuous representation of Virasoro algebra from the SSD Hamiltonian~\cite{Ishibashi:2016bey}.  In their representation, a generator $\mathcal{L}_{\kappa}$ is labeled by a continuous index $\kappa$ rather than an integer.   The origin of such continuous spectrum has been identified with the infinite range of the spatial coordinate on the complex plane.   Due to unconventional time development caused by the SSD Hamiltonian, equal time contours have a fixed point at $z=1$.  A parameter along each contour is no more  angular but extends to infinite range.  

Inspired by their result, the SSD technique has been applied to Takahashi-Tanimoto (TT) identity based solutions~\cite{Takahashi:2002ez}, which belong to universal sector of OSFT.   It has been known that a gauge fixed kinetic operator derived from a TT solution shares similar structure with  SSD Hamiltonian~\cite{Takahashi:2003xe}.  By applying the SSD technique, authors of~\cite{Kishimoto:2018ekq,Zeze:2018lao} have derived a continuous Virasoro generator $\mathcal{L}_{\kappa}$ for the TT tachyon solution. In this context, the continuous spectrum of Virasoro generator has a physical interpretation.  An open string loses its endpoints due to tachyon condensation since there are no more D-branes to attach with. A worldsheet for such string exhibits nontrivial geometry in which a parameter along a string is infinite as similar with the SSD case.   Thus the continuous spectrum is an important signal of tachyon condensation which distinguishes it from other backgrounds.   

On the other hand, the SSD-based technique has not yet been examined for wedge-based solutions~\cite{Schnabl:2005gv,Okawa:2006vm,Erler:2009uj} because of their rather different geometry.  A solution is not associated with single worldsheet but is rather a superposition of worldsheet strips with different width. No explicit information for lost boundaries of tachyon condensation is available.  The differences between identity based and wedge based frameworks can be simply explained by the different kinds of coordinate spaces.  In identity based framework, a solution is described by a scalar function $f(z)$ on a worldsheet with coordinate $z$.  On the other hand, in wedge bases framework a solution is described by a scalar function $G(K)$ on $K$-space, where $K$ generates a width of a worldsheet strip in sliver frame.  Therefore two frameworks can be regarded as $z$-space or $K$-space representation of open string fields.  

Aim of this paper is to obtain an unified understanding of geometric aspects common between two frameworks.  We accomplish this partially by introducing a Virasoro-like symmetry into wedge based framework.  We make use of the derivation string field $L$~\cite{Erler:2009uj,Mertes:2016vos} in $K B c $ algebra~\cite{Okawa:2006vm}.   Its action on a function of $K$ is
\begin{equation}
    [L, f(K)] = K \partial f(K),
\end{equation}
thus defines a vector field on $K$-space. It is straightforward to generalize this to Virasoro (Witt) generators on $K$-space by
\begin{equation}
 \mathbb{L}_{m} = - K^m L = - K^{m+1} \partial_{K}. 
\end{equation}
It is obvious that $\{\mathbb{L}_{m}\}$ obey Virasoro (Witt) algebra:
\begin{equation}
  [\mathbb{L}_{m}, \mathbb{L}_{n}]  =  (m-n)  \mathbb{L}_{m+n}. 
\end{equation}
Furthermore, as we will see later, a transformation generated by $\{\mathbb{L}_{m}\}$ leaves OSFT action invariant.  Remember that ordinary Virasoro algebra also generates a global symmetry of OSFT.  Therefore,  two kinds of generators provides an isomorphism of symmetries between $z$- and $K$-spaces.  

Next, we will seek for an analog of a deformed Virasoro generator found in identity based framework~\cite{Kishimoto:2018ekq,Zeze:2018lao}.  We successfully construct deformed Virasoro generators for the formal pure gauge solution~\cite{Okawa:2006vm}.   We find that the analytic structure of Virasoro generators around $K=0$ depends on a choice of a background.  To see this, let a deformed generator be  $\hat{\mathbb{L}}_{\lambda} $.  The $\lambda$ dependence of a generator is described a factor $\phi^{\lambda}$ which multiplies a generator.  For the perturvative vacuum, it behaves as $\phi^\lambda \sim K^\lambda $ therefore develops branch cuts around $K=0$ for non-integer $\lambda$.    For the tachyon vacuum, we find  $\phi^{\lambda} \sim (1+K)^\lambda$ which develops no singularities around the origin.  We also examine two-branes and find $\phi^{\lambda} \sim K^{\lambda} e^{-\frac{\lambda}{K}}$, which shares the  $K^{\lambda}$ factor with perturvative vacuum.   The behaviour of eigenfunctions observed here  somewhat resembles that of $z$-space eigenfunctions~\cite{Kishimoto:2018ekq,Zeze:2018lao}, therefore implies similarity between $z$- and $K$- spaces. 

This paper is organized as follows.  Section 2 provides basic framework used in this paper.  The $K B c L $ algebra is used to define a Virasoro generator $\mathbb{L}_{m}$ and a antighost mode $\mathbbm{b}_{m}$. It is shown that the Virasoro algebra generates global symmetry that commutes with the BRST charge $Q_{B}$.   A finite transformation is derived in section 3.  In section 4, we extend a global symmetry to nontrivial background described by the formal pure gauge solution~\cite{Okawa:2006vm}.  We study explicit examples of  the tachyon vacuum, perturvative vacuum (single brane) and two-branes.  We conclude in section 5.

\section{Virasoro algebra in $K$-space}
\subsection{Construction of Virasoro algebra}
Our main concern in this paper is a subspace of string fields spanned by four elements $K$, $B$, $c$ and $L$.  Former three are familiar ones~\cite{Okawa:2006vm} which have been used to construct various analytic solutions. A less familiar string field $L$ is defined by a left half integral of the energy-momentum tensor~\cite{Erler:2009uj,Mertes:2016vos}:
\begin{equation}
    L = \int_{C}  \frac{d z}{2 \pi i}
     \left(1+z^2
     \right) \left(    \arctan z   - \frac{\pi}{4}     \right)
     T (z) \ket{I},
\end{equation}
where $\ket{I}$ is the identity string field and $C$ is the right half of the unit circle ( $z= e^{i \theta},\ - \pi /2 \leq  \theta \leq \pi/2) $ and  integral runs counterclockwise.   $L$ is nothing but the familiar string field derivation through the relation
\begin{equation}
 [L, \Psi] = \frac{1}{2} (\mathcal{L}_{0}  -  \mathcal{L}_{0}^{\dagger}) \Psi. \label{familiarL}
\end{equation}
The derivation property of $L$ is manifest once written in an adjoint representation in the OSFT algebra (left hand side of \eqref{familiarL}) since
\begin{equation}
[L,  \Psi_{1}\Psi_{2}] = [L,  \Psi_{1}] \Psi_{2} 
+ \Psi_{1} [L,  \Psi_{2}]
\end{equation}
holds for any string fields $\Psi_{1}$ and $\Psi_{2}$.   Following Mertes and Schnabl, we extend $K B c$ algebra to $K B c L$~\cite{Mertes:2016vos}
\footnote{We do not include $B'$ of $K L B B' c$ algebra~\cite{Mertes:2016vos} since it will not appear in our analysis.}
\begin{equation}
Q_{B} c = c K c, \quad Q_{B} K = 0, \quad Q_{B} B =K, \quad Q_{B} L =0, \label{QonKBc}
\end{equation}
\begin{equation}
 c B + B c = 1, \quad c^2 =0,  
\end{equation}
\begin{equation}
[L, K] = K,  \quad  [L, B] = B, \quad [L, c] = -c. \label{weights}
\end{equation}

Let us proceed construction of  Virasoro generators in $K$-space.  As already explained in introduction,  first equation in~\eqref{weights} applies to a function of $K$ as
\begin{equation}
    [L, f(K)] = K \partial f (K), \label{Laction}
\end{equation}
where $f(K)$ is defined by a formal Laurent series
\begin{equation}
 f (K)  = \sum_{m} f_{m} K^m.
\end{equation}
According to \eqref{Laction}, $L$ can be regarded as a vector field on $K$-space.  Note that this vector field can be identified with $l_{0}$ of Witt algebra on the complex plane by replacing $K$ with $z$:
\begin{equation}
    l_{0} = - z \partial_{z} \quad
    \leftrightarrow  \quad 
    \mathbb{L}_{0}
    = - K \partial_{K}.
\end{equation}
It is natural to extend the $K$-space vector field to nonzero modes by\footnote{We can introduce another generator with different ordering of $K$ and $L$ by $
    l^{(p)}_{m} = - 
    K^{p} L K^{m-p}$, where $0 \leq  p \leq m $ is an integer.  We fix $p=m$ throughout this paper since such choice will not affect our analysis. 
}
\begin{equation}
\mathbb{L}_{m} = - K^{m+1} \partial_{K} 
 = - K^{m} L. \label{lmdef}
\end{equation}
The last expression of~\eqref{lmdef} defines a ``Virasoro generator'' in the space of string fields.  Using~\eqref{Laction}, one can derive $K$-space Virasoro algebra with zero central charge
\begin{equation}
    [\mathbb{L}_{m}, \mathbb{L}_{n}  ] = (m-n)
    \mathbb{L}_{m+n}.
\end{equation}

Here we would like to comment on difference between the string field representation of the Virasoro algebra introduced by Mertes and Schnabl \cite{Mertes:2016vos} and ours.  A major difference is that the former is linear with respect to the energy moment tensor frame while the latter is not. Another difference is that the Mertes-Schnabl algebra is a representation of conformal symmetry on $\tilde{z}$ plane.  On the other hand, our algebra on $K$-space is not a representation of conformal symmetry on the worldsheet.  

In addition to the Virasoro generators, we introduce an antighost mode by
\begin{equation}
    \mathbbm{b}_{m}
    = - B K^{m-1} L.
\end{equation}
Using \eqref{QonKBc} and \eqref{weights}, one can derive 
\begin{equation}
    Q_B  \mathbbm{b}_{m} 
    = \mathbb{L}_{m}, \quad
    [\mathbb{L}_{m}, \mathbbm{b}_{n}]
    = (m-n) \mathbbm{b}_{m+n}, \quad
    \{\mathbbm{b}_{m}, \mathbbm{b}_{n}\} =0,
\end{equation}
which are exactly identical to the algebra for conformal modes $L_{m}$ and $b_{m}$.

\subsection{Virasoro algebra as a global symmetry}

Next we will show that $\mathbb{L}_{m}$ generates a global symmetry of OSFT. We begin with a general discussion about a ghost number zero string field $J$.  We require $J$ to be $Q_{B}$ closed (i.e. $Q_B
 J = 0$).  It generates a  finite transformation on a string field $\Psi$: 
 \begin{equation}
     \begin{split}
    \Psi_{g} & = e^{J} \Psi e^{-J}  \label{Jtrans}\\
    & = \Psi + [J, \Psi]
 + \frac{1}{2!}[J,[J,\Psi]
 ]+ \cdots
\end{split}
 \end{equation}
where we expand exponentials in second line. We will show that~\eqref{Jtrans} defines an exact symmetry of OSFT action. For this purpose, the trace notation of the inner product between string fields is useful.  In this notation, OSFT action $S[\Psi]$ can be written as a sum of traces
\begin{equation}
    S [\Psi] 
    = \frac{1}{2} \mathrm{Tr}
    [\Psi Q_B \Psi] + \frac{1}{3}
    \mathrm{Tr} [ \Psi^{3}  ].
\end{equation}
Using cyclic property of trace and closedness of $J$, it is straightforward to confirm $S[\Psi_{g}] = S[\Psi]$.  This kind of symmetry has been called ``global symmetry'' in literature according to the formal correspondence between OSFT and Chern-Simons theory where $Q_B$ is identified with the exterior derivative $d$~\cite{Witten:1985cc}.

Now let us apply above discussion to Virasoro generators by setting $J \sim \mathbb{L}_{m}$.  $Q_B$-closedness of $\mathbb{L}_{m}$ is obvious since it is a product of  $Q_{B}$-closed fields $K$ and $L$. Thus any linear combination of Virasoro generators can be regarded as a generator of a global symmetry of OSFT.  A linear combination of generators can be specified by coefficients $\{v_m\}$  therefore introduces a vector field on $K$-space.
\begin{equation}
    \begin{split}
    \mathbb{L}_{v} & =
    \sum_{m} v_{m} 
    \mathbb{L}_{m} \\ 
    & = 
   - \sum v_{m} K^{m} L \\
    &= -v(K) \partial_{K},
\end{split}
\label{lvdef}
\end{equation}
where we set $v(K) = \sum_{m} v_{m} K^{m +1} $ in the third line.  

We again would like to make a comparison between global symmetries in $z$- and $K$-spaces.   The former is described by a contour integral of the energy momentum tensor in $z$-space
\begin{equation}
    L_{v} = \oint \frac{dz}{2 \pi i}\, v(z) T(z),
    \label{Jcont}
\end{equation}
where $v(z)$ is restricted to a midpoint-preserving vector field~\cite{Witten:1985cc}
\begin{equation}
    v(z) = \sum_{m} v_{m} (z^{m+1} 
    - (-1)^{m} z^{-m+1}   )
    \label{vrest}
\end{equation}
to make cubic term $\mathrm{Tr} [\Psi^3]$ invariant under an infinitesimal transformation. This symmetry is a consequence of conservation law on the worldsheet, which allows us to deform the closed contour in~\eqref{Jcont} to different radius. On the other hand, the $K$-space generator~\eqref{lvdef} does not allow such interpretation. In addition, coefficients $v_{m}$ in~\eqref{lvdef} is not restricted like~\eqref{vrest} but can be chosen arbitrary.   

\section{Finite transformation}
Next we evaluate finite transformations generated by $K$-space Virasoro generators.  Here we consider a finite transformation generated by $\mathbb{L}_{v}$ of~\eqref{lvdef} and its action onto $K B c L$ elements.  First, we evaluate a finite transformation of $K$
\begin{equation}
    e^{\mathbb{L}_{v}} K e^{-\mathbb{L}_{v}}.
    \label{lvfinite}
\end{equation}
Remember that $\mathbb{L}_{v}$ coincides with a conventional expression of a Witt generator $-v(z)\partial_{z} $  by simply replacing $K$ with $z$.  Therefore, we can translate the standard formula of global conformal transformation on $z$-space to $K$-space. On $z$-space, a finite conformal transformation $f(z)$ is evaluated by the formula~\cite{LeClair:1988sp,Schnabl:2002gg}
\begin{equation}
   f(z) = e^{-v(z) \partial_{z} }
    z e^{v(z) \partial_{z}} = e^{-v(z) \partial_{z} } 
    z . \label{julia}
\end{equation}
Here we only need to do exactly same thing as~\eqref{lvfinite} for $K$.  Thus we obtain
\begin{equation}
     e^{\mathbb{L}_{v}} K e^{-\mathbb{L}_{v} } = f(K).
     \label{lvfinitecomp}
\end{equation} 
where $v(K) = \sum_{n} v_{n} K^{n+1}$ and $f(K) = e^{-v(K) \partial} K$.  

Next, finite transformation of $B$ is obtained just by ``integrating'' \eqref{lvfinitecomp} with respect to $Q_{B}$:
\begin{equation}
     e^{\mathbb{L}_{v}} B  e^{-\mathbb{L}_{v} }= \frac{f(K)}{K} B.
\end{equation}
Finite transformation of $c$ is somewhat complicated as it is not closed within $K B c $ space.  To see this, we evaluate infinitesimal transformation of $c$: 
\begin{equation}
    [\mathbb{L}_v, c]
    = \frac{v}{K} c - \left[\frac{v}{K},c\right] L. \label{Lmc}
\end{equation}
Note that the right hand side involves a term proportional to $L$ .  Repeated application of ~\eqref{Lmc} in finite transformation involves higher powers of $L$.  As a result, final expression of the finite transformation becomes complicated.  Therefore we just write it as
\begin{equation}
    e^{\mathbb{L}_{v}} c  e^{-\mathbb{L}_{v} }
    = e^{\mathrm{ad}_{\mathbb{L}_{v}} } c.
\end{equation}
Finally, finite transformation of $L$ can be derived by writing $L = -\mathbb{L}_0$ and applying finite transformation law, or writing $L = - K/v(K) \cdot \mathbb{L}_{v}$ and applying a finite transformation only on the first factor. In either way one obtain
\begin{equation}
    e^{\mathbb{L}_{v}} L e^{-\mathbb{L}_{v} }
    = \frac{f}{K \partial f} L.  \label{Ltransf}
 \end{equation}

Here we would like to apply a global transformation to a formal pure-gauge solution of the form~\cite{Okawa:2006vm}
\begin{equation}
    \Psi_{G} = \sqrt{1-G} c \frac{K}{G} B c 
    \sqrt{1-G}, \label{okawa}
\end{equation}
where $G = G(K)$ is a function of $K$.   Here we quote finite transformations on $K B c$ elements again
\begin{equation}
    K'  = f,\quad
    B'  = \frac{f}{K} B,\quad 
    c_{1}'  =  e^{\mathrm{ad}_{\mathbb{L}_{v}} } c.
    \label{KBcglobal}
\end{equation}Applying these to \eqref{okawa} yields
\begin{equation}
    \Psi_{G}'  = \sqrt{1-G'}
    c'_{1} \frac{f^2}{G'} B c'_{1}
    \sqrt{1-G'} \label{okawalike}
\end{equation}
where $G'(K) = G(K')$.  This is an equivalent representation to the original solution~\eqref{okawa} as long as a transformation is regular.  Values of gauge invariant quantities remain unchanged under this transformation.   Note that a transformed solution does not preserve the original structure of~\eqref{okawa} due to the $L$ dependence of $c'_{1}$.  

Interestingly, there exists another transformation which preserves the original structure of~\eqref{okawa}:\cite{Erler:2012dz,Masuda:2012kt,Hata:2012cy}
\begin{equation}
    K'  = f,\quad
    B'  = \frac{f}{K} B,\quad 
    c_{2}'  =   c\frac{K}{f} B c.
    \label{emnt}
\end{equation}
This has been known as EMNT transformation~\cite{Erler:2012dz,Masuda:2012kt,Hata:2012cy}.  One can easily apply~\eqref{emnt} to the formal pure gauge solution and confirm that it only affects on $G$:  
\begin{equation}
\Psi_{G}'  = \sqrt{1-G'}
    c \frac{K}{G'} B c
    \sqrt{1-G'}.
\end{equation}
 The transformations of~\eqref{emnt} are also global since they commute with $Q_{B}$.  It also should be noted that first two transformations of~\eqref{KBcglobal} and~\eqref{emnt} are identical.   Unfortunately, we have not yet identified an infinitesimal generator of EMNT transformation within $K B c L$ space.

\section{Virasoro algebra for nontrivial backgrounds}

\subsection{Deformed Virasoro generators}
We have derived Virasoro algebra as a global symmetry of OSFT described by $Q_B$.  Let us extend it to other background described by a deformed BRST charge
\begin{equation}
 \hat{Q} \Psi =  Q_{B} \Psi 
 + \Psi_{G} \Psi + \Psi \Psi_{G}
 \label{Qhat}
\end{equation}
where $\Psi_{G}$ is given by~\eqref{okawa}. We first note how the original global symmetry must be broken by the shift of a background.   It can be shown that any choice of $G$ other than a perturbative vacuum $G = \alpha K$ breaks the global symmetry completely.  Therefore, if a global symmetry exists on a deformed background, it should be described by generators other than $\mathbb{L}_{m}$.   

Here we employ the strategy of the Sine Square Deformation (SSD) based studies~\cite{Kishimoto:2018ekq,Zeze:2018lao,Ishibashi:2016bey}. 
An outlined of the strategy is
\begin{enumerate}
    \item Define a deformed Hamiltonian
    \item Derive an eigenmodes of a Hamiltonian
    \item Use obtained eigenmodes to derive Virasoro generators and confirm Virasoro algebra
\end{enumerate}

Let us perform each step following with the strategy.   In~\cite{Kishimoto:2018ekq,Zeze:2018lao}, first step has been carried out by simply applying the deformed BRST charge to the antighost zero mode.
We do same thing by applying the BRST charge \eqref{Qhat} to  the ghost zero mode $\mathbbm{b}_0$:
\begin{align}
\hat{\mathbb{L}}_{0}
 & = \hat{Q}  \mathbbm{b}_0 \\
 & = u_{0} + v_{0} L,
\end{align}
here $u_{0}$ and $v_{0}$ are $L$ independent string fields.  Explicit form of them can be derived from~\eqref{Qhat}: 
\begin{align}
     u_{0} & =
     -  \frac{K \partial F}{G} B c F
     - \frac{ F}{G} B c K \partial F
     - K F \partial \left(\frac{1}{G}\right) B c F,
     \\
v_{0} & = 
-\frac{1}{G}
    -  \frac{F}{G}  B c F
    +  F B c \frac{ F}{G},
\end{align}
where $F=\sqrt{1-G}$.   

Second step is identification of eigenmodes of $\hat{\mathbb{L}}_{0}$.  We begin with an ansatz for an antighost eigenmode
\begin{equation}
    \hat{\mathbbm{b}}_{\lambda}
    =B \beta_{\lambda} L
\end{equation}
where $\beta_{\lambda}$ is a function of $K$ with an eigenvalue $\lambda$.  We then solve an eigenvalue equation
\begin{equation}
    [\hat{\mathbb{L}}_{0},  
     \hat{\mathbbm{b}}_{\lambda}] = \lambda
     \hat{\mathbbm{b}}_{\lambda}.\label{beq}
\end{equation} After a little algebra with~\eqref{beq}, we obtain a differential equation
\begin{equation}
     \frac{
    \beta_{\lambda} + K \partial \beta_{\lambda} }{G} 
    = \lambda \beta_{\lambda}. \label{diffeq}
\end{equation}
By integrating this equation we find
\begin{equation}
    \beta_{\lambda} = - \frac{\phi^{\lambda}}{K}, \label{beta}
\end{equation}
where
\begin{equation}
    \phi (K) = \exp \left(\int^{K} d K' \frac{ G(K') }{K'}
    \right). \label{phi}
\end{equation}
Here we have chosen an integration constant so that $\mathbbm{b}_{\lambda}$ reduces
to $\mathbbm{b}_{m}$ for a trivial choice $G(K) = 1$.  As a result, an antigost mode turned out to be
\begin{equation}
    \hat{\mathbbm{b}}_{\lambda} = - B  \frac{\phi^{\lambda} }{ K}  L.
    \label{bsolution}
\end{equation}
Once an antighost mode is obtained, third step is straightforward. A Virasoro generator is obtained by operating deformed BRST charge onto it.
\begin{equation}
    \begin{split}
    \hat{\mathbb{L}}_{\lambda}
    & =  \hat{Q} \hat{\mathbbm{b}}_{\lambda}  \\
    & = u_{\lambda} + v_{\lambda} L, 
\end{split}
\label{lhatresult}
\end{equation}

where
\begin{align}
u_{\lambda} & = \phi^{\lambda} u_{0}, \label{u}\\
v_{\lambda} & = -\frac{\phi^{\lambda}}{G}
-  \phi^{\lambda}\frac{F }{G}  B c F 
-  F B c \frac{F}{G} \phi^{\lambda}. \label{v}
\end{align}
It is obvious that $\hat{\mathbb{L}}_{\lambda}$  generates a global symmetry of OSFT defined by $\hat{Q}$ since it is manifestly $\hat{Q}$ exact.  Furthermore, we can show that the deformed generators obey Virasoro algebra
\begin{equation}
[ \hat{ \mathbb{L}}_{\lambda},  \hat{\mathbb{L}}_{\lambda'} ]
= (\lambda -\lambda') \label{deformedvirasoro}
\hat{\mathbb{L}}_{\lambda+ \lambda'}.
\end{equation}One can derive this directly with the expressions~\eqref{lhatresult}, \eqref{u}, \eqref{v} and \eqref{phi}.  Alternatively, it is rather easier to show 
\begin{equation}
    [\hat{\mathbbm{b}}_{\lambda}, \hat{\mathbb{L}}_{\lambda'}   ]
    = (\lambda - \lambda') \hat{\mathbbm{b}}_{\lambda+\lambda'}
    \label{bleq}
\end{equation} 
first and multiply the whole equation by $\hat{Q}$ to obtain~\eqref{deformedvirasoro}.   We show the derivation of~\eqref{bleq} in appendix. 

\subsection{Examples}

As we have seen in previous section, eigenmodes  $\hat{\mathbb{L}}_{\lambda}$ and $\hat{\mathbbm{b}_{\lambda}}$ are characterized by a function $\phi^{\lambda}$ of~\eqref{phi}.  Let us evaluate this function explicitly for known solutions.  Here we consider $G(K)$ for multibranes~\cite{Murata:2011ep,Hata:2011ke}.  
\begin{equation}
    G(K) = \left(\frac{1+K}{K} \right)^{n}, \label{Gn}
\end{equation}
where we restrict $n$ to be $n =-1$ (the tachyon vacuum), $n=0$ (perturvative vacuum or one brane), $n=1$ (two branes) since  they are known as consistent solutions that 
can reproduce expected values of gauge invariant observables~\cite{Murata:2011ep,Hata:2011ke}
\footnote{Solutions for other values of $n$ have been proposed by Hata~\cite{Hata:2019dwu}.  Our analysis can be applied to them.  We leave it as a future task.}.
Then explicit form of $\phi^{\lambda}$ can be obtained by performing integration in \eqref{phi}.  Fortunately, the integral can be preformed analytically for all of three examples.  The results are 
 \begin{equation}
 \phi^{\lambda} (K) \sim  
 \begin{cases}
 (1+K)^{\lambda} &\text{for the tachyon vacuum}\ (n=-1) \\
 K^{\lambda}&  \text{for  perturbative vacuum}\ (n=0)\\
  e^{-\frac{\lambda}{K}} K^{\lambda} &  \text{for two-branes}\ (n=1)
 \end{cases}\label{kspacesol}
 \end{equation}

Let us compare these with the $z$-plane results for identity based solutions.  In that setting, a deformed Virasoro generator is given by
\begin{equation}
\mathcal{L}_{\lambda} 
= \int_{C_{+}} \frac{dz}{2 \pi i}
g(z) f_{\lambda} (z)  \mathcal{T} (z),
\end{equation}
where $C_{+}$ is the upper half of the unit circle and $\mathcal{T}(z)$ is the framed version of the twisted energy momentum tensor~\cite{Kishimoto:2018ekq}.  A function $f_{\lambda} (z)$ determines the $\lambda$ dependence of a generator therefore plays a role analogous to $\phi^{\lambda} (K)$ .   Examples of $f_{\lambda}(z)$ to be compared with
our result are 
 \begin{equation}
 f_{\lambda} (z) \sim  
 \begin{cases}
 e^{\frac{2 \lambda}{z^2 -1}} &  \text{for the tachyon vacuum}\\
 z^{\lambda} & \text{for the perturbative vacuum} 
 \end{cases}\label{zspacesol}
 \end{equation}
 We find similarity between the perturbative vacuum results~\eqref{kspacesol} and~\eqref{zspacesol} as both are $\lambda$th power of variables $K$ or $z$.   On the other hand, results for the tachyon vacuum are not quite same between two frameworks.  However, they share similar feature that a potential singularities around the origin ($K=0$ or $z=0$) seen in the perturbative vacuum disappear in the tachyon vacuum.    For two-branes we cannot make a comparison between two frameworks since two-brane solution has not yet been known in the identity based framework.   
 
 Next we pay attention to the $K$-space results~\eqref{kspacesol} for one and two branes, with a expectation that they share similar feature since both have nontrivial cohomologies.  Indeed they share common factor of $K^\lambda$, although two branes exhibits an essential singularity. 
 
 In identity based formalism, $\lambda$ dependence of a function $f_{\lambda} (z)$ determines the spectrum~\cite{Kishimoto:2018ekq,Zeze:2018lao}. On perturvative vacuum, requirement for single-valuedness of $f_{\lambda} (z)$ restricts $\lambda$ to an integer. On the other hand,  the tachyon vacuum admits non-integer value of $\lambda$.   The requirement for single valuedness of $f_{\lambda} (z) $ is reasonable since it defines conserved quantities of conformal field theory on $z$ plane.    
 
 On the other hand, we are not quite sure whether the $\lambda$ dependence of $\phi(K)^\lambda$ determines a spectrum as similar with the $z$ plane analysis.   It is merely because we do not know which conditions should be imposed on a function on $K$-space.  One possible condition will come from requirement for geometric expression 
 \begin{equation}
 \phi^{\lambda} (K) = \int_{0}^{\infty} dt
 \Phi_{\lambda} (t) e^{-t K}. \label{laplace2}
 \end{equation}
where $\Phi_{\lambda} (t)$ is an inverse Laplace transform of $\phi^\lambda (K)$.  However,  it seems that $\lambda$ need not to be restricted to an integer on perturbative vacuum. In fact, the inverse Laplace transform $\Phi_{\lambda} (t)$ can be defined for an eigenfunction $\phi^{\lambda} \sim K^{\lambda}$ even for non-integer $\lambda$.  For negative $\lambda$, the equation~\eqref{laplace2} can be definitely written
\begin{equation}
 \frac{1}{K^\nu} = \int_{0}^{\infty} dt 
 \frac{t^\nu}{\Gamma(\nu)} e^{-t K},
\end{equation}
 where we set $\nu = - \lambda$. Even for positive $\lambda$,  we could find $\Phi_{\lambda} (t)$ as a distribution rather than ordinary function. 

\section{Conclusions}
We have introduced $K$-space Virasoro generators as elements of $K B c L$ algebra.  The string fields $\mathbb{L}_{m}$ and $\mathbbm{b}_{m}$ have been shown to obey same algebraic relations as that of conventional CFT. We  have shown that $\{ \mathbb{L}_{m}\}$ generate global symmetry of OSFT.  Explicit formulas for finite transformations of $K B c L$ elements are derived.  

Next, we have extended it to deformed open string backgrounds described by the Okawa formal pure gauge ansatz.  By employing the SSD based strategy, we have derived explicit form of the Virasoro generators.  We have shown that they obeys Virasoro algebra and are generators of a global symmetry around a deformed background.

We have investigated examples for 0, 1 and 2 D-branes by choosing a function $G(K)$ correspondingly.   We have observed that the eigenvalue dependence of the generators somewhat resembles those derived for identity based solutions.  

We conclude that all of our results implies similarity between $z$-space (identity based framework) and $K$-space (wedge based framework), because we have found global symmetries that share same algebra and similar behavior of Virasoro generators against changes of background.   Given these results, one may ask whether $K$ plays fundamental role as $z$ does in CFT.  To answer this question, we have to understand more aspects of $K$-space.   For example, the splitting into left and right moving sectors observed in identity based framework~\cite{Kishimoto:2018ekq} remains unexplained in wedge based framework.   In addition, we do not yet have an interpretation of the imaginary direction in $K$-space.   

\begin{appendices}
\section{Proof of \eqref{bleq}}

First we split the commutator into three terms:
\begin{align}
[\hat{\mathbbm{b}}_{\lambda},  \hat{\mathbb{L}}_{\lambda'} ] & =
[\hat{\mathbbm{b}}_{\lambda},  u_{\lambda'} + v_{\lambda'} L ] \\
& = [\hat{\mathbbm{b}}_{\lambda}, u_{\lambda'} ]
+ [\hat{\mathbbm{b}}_{\lambda}, v_{\lambda'} ] L
+ v_{\lambda'} [ \hat{\mathbbm{b}}_{\lambda}, L]. \label{threeterms}  
\end{align}
Each term can be calculated separately.  First term of~\eqref{threeterms} becomes
\begin{equation}
    \begin{split}
 [\hat{\mathbbm{b}}_{\lambda}, u_{\lambda'} ] & = -\frac{2 K F \partial F}{G^2} B \beta_{\lambda} \beta_{\lambda}' L \\
 & = \frac{2 K F \partial F}{G^2} 
 \hat{\mathbbm{b}}_{\lambda+ \lambda'}. \label{A1}
\end{split}
\end{equation}
In second line we used the fact that the solution~\eqref{beta} satisfies $\beta_{\lambda}\beta_{\lambda'} = \beta_{\lambda + \lambda'}$.    

Second term of~\eqref{threeterms} simplifies since last two terms in $v_{\lambda'}$ cancel each other
in the commutator:
\begin{equation}
\begin{split}
[\hat{\mathbbm{b}}_{\lambda}, v_{\lambda'} ] L & =
\left[\hat{\mathbbm{b}}_{\lambda}, \frac{K \beta_{\lambda'}}{F}
- F B c F \frac{K \beta_{\lambda'}}{G} +
\frac{K \beta_{\lambda'}}{G} F B c F \right] L \\
& = \left[\hat{\mathbbm{b}}_{\lambda}, \frac{K \beta_{\lambda'}}{F}
\right]L.
\end{split}
\end{equation}
The commutator in front of $L$ involves derivative of $\beta_{\lambda'}$, which can be transformed into a quantity without derivative by making use of~\eqref{diffeq}.  We obtain
\begin{equation}
[\hat{\mathbbm{b}}_{\lambda}, v_{\lambda'} ] L
= -\left(
 \lambda' + \frac{2 K F \partial F}{G^2} 
\right) \hat{\mathbbm{b}}_{\lambda+ \lambda'}. \label{A2}
\end{equation}Third term of~\eqref{threeterms} is represented as
\begin{equation}
v_{\lambda'} [ \hat{\mathbbm{b}}_{\lambda}, L]
=  v_{\lambda'} [B \beta_{\lambda},L]L. \label{thirdbraket}
\end{equation}As similar with the evaluation of the second term, a commutator inside right hand side of the above equation involves derivative of $\beta_{\lambda}$.  Again by making use of~\eqref{diffeq} we obtain
\begin{equation}
[B \beta_{\lambda},L] = - \lambda B \beta_{\lambda} G.
\end{equation}Plugging this back to the right hand side of~\eqref{thirdbraket} we just have
\begin{equation}
v_{\lambda'} [ \hat{\mathbbm{b}}_{\lambda}, L] = \lambda \hat{\mathbbm{b}}_{\lambda+ \lambda'}. \label{A3}
\end{equation}Finally we assemble three terms~\eqref{A1},~\eqref{A2} and~\eqref{A3}.  We observe that the first term\eqref{A1}, which involves nontrivial $K$-dependent factor, is canceled with a part of~\eqref{A2}.   Thus we arrive at
\begin{equation}
[\hat{\mathbbm{b}}_{\lambda},  \hat{\mathbb{L}}_{\lambda'} ] 
= (\lambda-\lambda') \hat{\mathbbm{b}}_{\lambda+\lambda'}.
\end{equation}
\end{appendices}

\bibliographystyle{utphys}
\bibliography{hep}

\end{document}